%% file: AdaIN-BF_SK_final.tex
\documentclass[journal]{IEEEtran}

\usepackage{cite}

\input{packages}
\input{macros}

\begin{document}

\title{Switchable Deep Beamformer
}
\author{
Shujaat Khan,
        Jaeyoung Huh,~%\IEEEmembership{Student Member,~IEEE,}
%XXX, 
        and~Jong~Chul~Ye,~\IEEEmembership{Fellow,~IEEE}
\thanks{This work was supported by the National Research Foundation (NRF) of Korea grant NRF-2020R1A2B5B03001980. The authors are with the Department of Bio and Brain Engineering, Korea Advanced Institute of Science and Technology (KAIST), Daejeon 34141, Republic of Korea (e-mail:\{shujaat,woori93,jong.ye\}@kaist.ac.kr).}}% <-this % stops a space	

\maketitle

% As a general rule, do not put math, special symbols or citations in the abstract or keywords.
\begin{abstract}
Recent proposals of  deep beamformers for ultrasound imaging (US) using deep neural networks have attracted significant attention as computational efficient alternatives to adaptive and compressive beamformers.
Moreover, deep beamformers are versatile in that image post-processing algorithms can be combined with the beamforming.
Unfortunately, in the current  technology,
a separate beamformer should be trained and stored for each application, demanding significant scanner resources.
To address this problem, here we propose a {\em switchable} deep beamformer that can produce various types of output 
such as DAS, speckle removal, deconvolution, etc., using a  single network with a simple switch.
In particular, the switch is implemented through Adaptive Instance Normalization (AdaIN) layers, so that distinct outputs can be generated by merely
changing the AdaIN code. 
 Experimental results using B-mode focused ultrasound confirm the flexibility and efficacy of the proposed method for various applications.
\end{abstract}


\IEEEpeerreviewmaketitle

\section{Introduction}
\label{sec:introduction}

Ultrasound (US) is one of the most versatile medical imaging modalities. Thanks to its fast frame rate and radiation free nature, it is the first choice for  applications such as fetal imaging,  cardiac imaging, etc. 

The image generation process in US involves scanning and reconstruction steps.
In focused mode, the region of interest (ROI) is scanned in a line-by-line fashion, while in planewave 
(PW) mode,  the full ROI is scanned at once. 
For the reconstruction, out-of-phased RF signals are accumulated after back-propagation. For example, the basic steps of conventional delay-and-sum (DAS) beamforming pipeline are: (1) time-of-flight correction in which RF measurements are shifted in time, and (2) the summation of the time-corrected signal.

% In standard delay-and-sum (DAS) beamformer fixed-value are used as apodization weight. 
While the DAS beamformer is  widely used  due to its simplicity and robust reconstruction performance, 
it  often suffers from  poor resolution due to signal side-lobes.  Moreover, DAS images are usually associated with image artifacts such as speckles, blurring, etc., whose removal have been one of the main research topics in US research.
%Since the accuracy of the estimated time-of-flight is linked to the limited bandwidth of the transducer, therefore to generate a clinically acceptable quality image the DAS require large number of receivers (Rx).
However, existing approaches using adaptive beamforming techniques
%To resolve the aforementioned issues, variety of adaptive beamforming methods were developed 
\cite{AdaptiveBF1,CaponBF,CaponBF2,MVBF1,MVBF2,BSBF1,FastRobustBF,MultiBeamBF1,IterativeBF1}
and compressed sensing \cite{jin2016compressive, yoon2018efficient, burshtein2016sub, schiffner2011fast} are computationally expensive and sensitive to hyperparameters, so
a recent trend is using deep learning approaches in the design of beamformers and image processing algorithms  \cite{luchies2018deep, nair2018deep, hyun2019beamforming,khan2019deep,khan2020adaptive,luijten2020adaptive}.

%%Each method has its own merits and demerits. In adaptive beamformers different apodization weights are used for each channel. The weights are estimated using the array-statistics such as correlation, coherence and cohesion resulting in improved contrast and resolution. One of the widely studied adaptive beamforming method is Capon beamforming \cite{CaponBF,CaponBF2,MVBF1}. The apodiation weights in Capon beamformer are derived by minimum variance (MV) method which requires computationally expensive calculation of the inverse of spatial covariance matrix of channel data\cite{MVBF2}. To reduce the computational complexity of Capon (MV) beamformer, in recent past, number of improved solutions have been proposed \cite{MVBF1,MVBF2,BSBF1,FastRobustBF, MultiBeamBF1, IterativeBF1}.  In typical settings the MV beamformer shows better reconstruction performance than the standard DAS, however, it also require large number of measurements. The major bottleneck in conventional beamforming methods is that they are designed with ray approximation, whereas ultrasound signal exhibit many wave phenomenon such as reflection, reflection, and diffraction, etc. 
%
%To resolve aforementioned challenges of the conventional beamforming methods, recently inspired by tremendous success of deep learning in medical imaging number of novel solutions has been developed \cite{luchies2018deep, nair2018deep, hyun2019beamforming, luijten2020adaptive, khan2020adaptive}. 

\begin{figure*}[!hbt]
	%	\centerin
	\centerline{\epsfig{figure=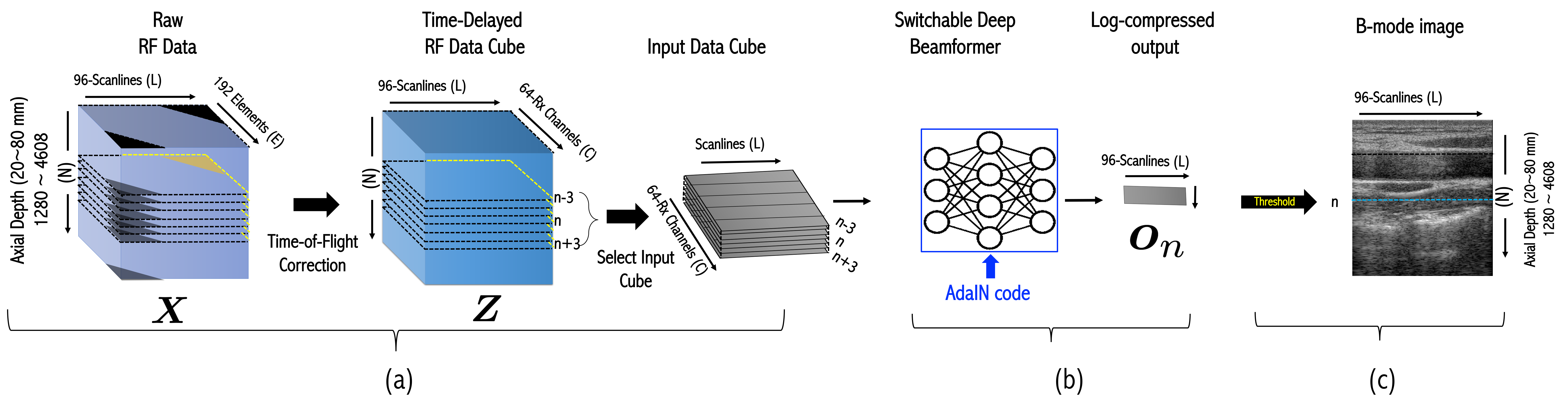, width=18cm}}   
	\caption{\footnotesize Data processing pipeline for the proposed switchable deep beamformer. (a) Neural network input preparation step,
	 (b) switchable deep beamformer, and (c) thresholding and display step for visualization.}
	\label{fig:data_processing_pipeline}	
\end{figure*}

For example,  Luchies and Byram  \cite{luchies2018deep} proposed a deep learning based frequency domain beamforming method for off-axis scattering suppression. % In \cite{nair2018deep}, using neural network an alternative of beamformer is designed to estimate the segmentation mask of a phantom from RF domain data. The models in \cite{luchies2018deep, nair2018deep} were trained to mimic desired map obtained using simulation data. 
In \cite{hyun2019beamforming},  a beamformer is designed  to produce speckle-free images from channel data. A major limitation of these methods \cite{luchies2018deep,hyun2019beamforming}  is that the training data are usually obtained from simulation phantom and therefore recovery of accurate images   is challenging especially for in-vivo data.
On the other hand,   recent deep beamformers \cite{khan2019deep,khan2020adaptive,luijten2020adaptive} utilizes  in vivo data as reference target for supervised training.
For example, in the universal deep beamformer \cite{khan2019deep,khan2020adaptive}, neural network is trained to generate  high resolution
DAS, minimum variance beamformer (MVBF), and deconvolution ultrasound outputs from  channel data.  In another work \cite{luijten2020adaptive}, a neural
network was trained to mimic the output of MV beamformer for fast beamforming.

Although deep beamfomers provide impressive
performance and ultra-fast reconstruction, one of the downsides of the deep beamformers is that
a distinct model is needed for each type of desired output. For instance, to obtained DAS output, a model is needed which mimics DAS; similarly, for MVBF a separate model is needed. Although the architecture of the model will be the same,  separate weight is needed to be stored  for each output type.  Additionally, to enable a deep beamformer to generate
output that matches to several post processing  results, such as deblurring or speckle reduction, etc.,
 it further increases the number of models to be stored.

To address this  issue,  here we propose a novel {\em switchable}
deep beamformer architecture using    adaptive instance normalization (AdaIN) layers.
AdaIN was originally proposed as an image style transfer method, 
in which the mean and variance of the feature vectors are replaced by those of the style reference image \cite{huang2017arbitrary}.
Recently, AdaIN has attracted significant attention as a powerful component to control generative models.
For example, in the StyleGAN \cite{karras2019style}, the network generated various realistic faces with the same inputs by simply adjusting the mean and variance of the feature maps using AdaIN.
In StarGANv2 \cite{choi2020stargan}, AdaIN plays the key role in converting styles from input images.
Furthermore, a recent study \cite{mroueh2019wasserstein} has shown that the change of style by AdaIN is not a cosmetic change but a real one thanks to the important link to the optimal transport \cite{peyre2019computational,villani2008optimal} between the feature vector spaces.

Inspired by the success of AdaIN and its theoretical understanding, 
one of the most important contributions of this work is to demonstrate that
a {\em single} deep beamformer with AdaIN layers can learn target images from various styles. % for training.
Here, a ``style'' refers to a specific output processing, such as DAS, MVBF, deconvolution image, despeckled images, etc.
Once the network is trained,
the deep beamformer can then generate various style output
by simply changing the AdaIN code.
Furthermore, the AdaIN code generation can be easily done with a very light  AdaIN code generator, 
so the additional memory overhead at the training step is minimal.
Once the neural network is trained,   we only need the AdaIN codes without the generator, 
which makes the system even simpler.
As such, our switchable deep beamformer can be
considered as the first, fully adaptive multi-purpose deep beamformer  for the reconstruction of high-quality medical ultrasound images.

This paper is  organized as follows.
The deep beamformer and AdaIN are first reviewed in Section~\ref{sec:review},
after which we explain how these techniques can be synergistically combined to produce
a switchable deep beamformer in Section~\ref{sec:theory}.
 Section~\ref{sec:methods} describes the data set and experimental setup. Experimental results are provided in Section~\ref{sec:results}, which is followed
by  Conclusions in Section~\ref{sec:conclusion}.

\begin{figure*}[!hbt]
	%	\centerin
	\centerline{\epsfig{figure=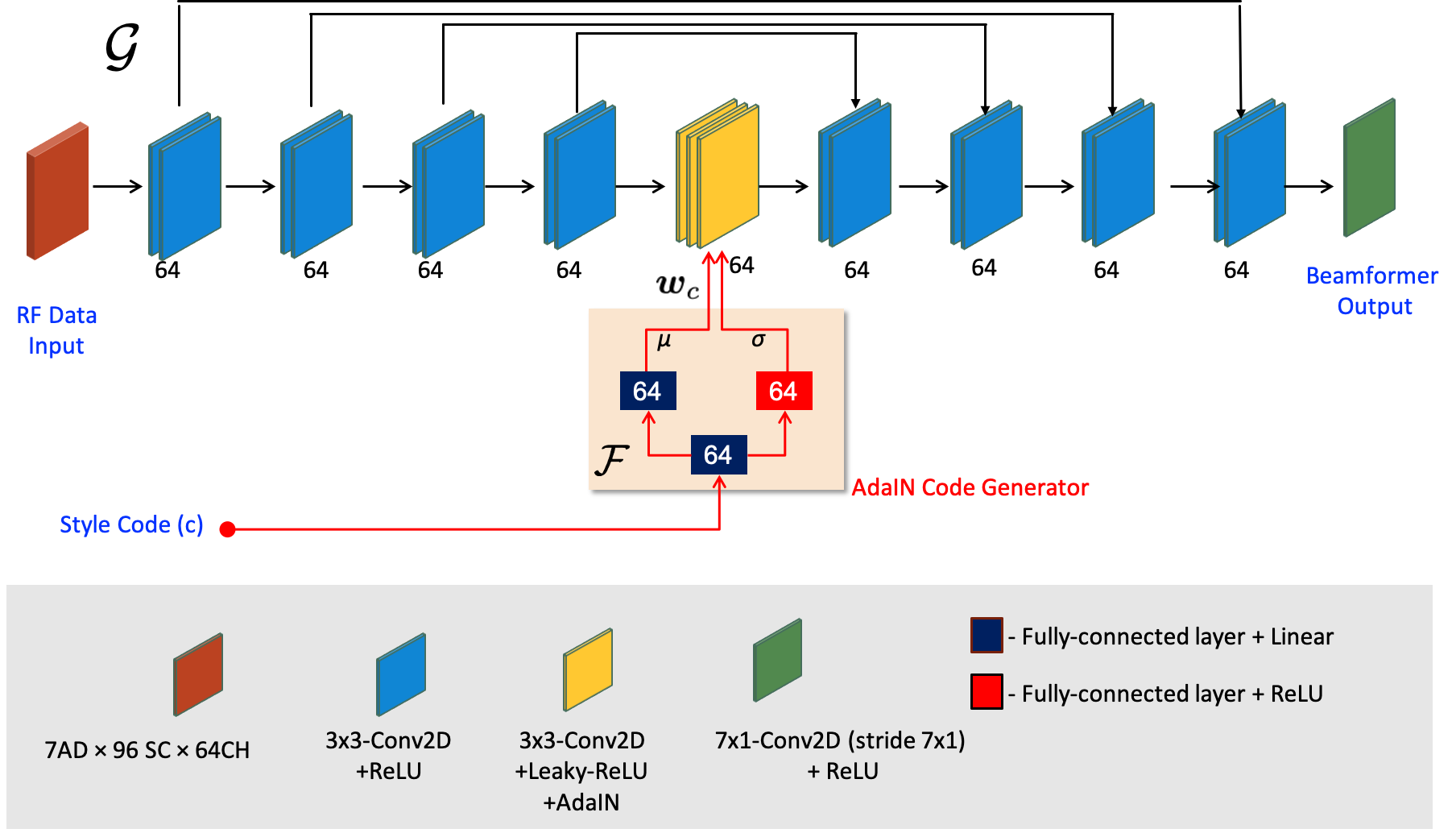, width=14cm}}   
	\caption{\footnotesize Architecture of our switchable deep beamformer.  Adaptive instance normalization (AdaIN) layer
	is added at the bottleneck layer, for which AdaIN code is generated by a separate AdaIN code generator network.}
	\label{fig:network_architecture}	
\end{figure*}

\section{Related works}
%\section{Mathematical Preliminaries}
\label{sec:review}

%{In this section, we review some mathematical preliminaries and existing beamforming methods.}

\subsection{Deep Beamformer}

Among the various forms of deep beamformers, here we review the universal deep beamformer  (DeepBF) \cite{khan2019deep,khan2020adaptive} as a representative form of deep beamformer, since  our switchable deep beamformer is built upon this architecture.

In US, measured RF data  is  usually given as a three-dimensional cube $\Xb \in \Rd^{L\times N\times E}$  as shown in Fig.~\ref{fig:data_processing_pipeline}(a),
where $L, N,$ and $E$ denote the number of scan lines (or transmit events (TE)), depth planes, and the number of probe elements, respectively.
%The RF data cube $\Xb$ is often represented as $\Xb:=[\xb_{l,n}]_{l,n}$, where
%$\xb_{l,n}\in \Rd^{E}$ is the $(l,n)$-th element of the data cube, representing  the RF data measured by the receiver channels  from the the $l$-th scan line at the depth index $n$.
The time-delay corrected data cube $\Yb \in \Rd^{L\times N\times E}$ is similarly denoted by
$\Yb=[\yb_{l,n}]_{l,n}$, where
 \begin{eqnarray}
 \yb_{l,n}=\begin{bmatrix} y_{l,n}[0] & y_{l,n}[1] & \cdots &y_{l,n}[E-1]\end{bmatrix}^\top \in \Rd^{E}
 \end{eqnarray}
Then,  the received channel data can be explicitly modeled as
 $\Zb=[\zb_{l,n}]_{l,n}$, where
 \begin{eqnarray}
 \zb_{l,n}=\begin{bmatrix} z_{l,n}[0] & z_{l,n}[1] & \cdots &z_{l,n}[{J}-1]\end{bmatrix}^\top \in \Rd^{J}
 \end{eqnarray}
and
 \begin{align}\label{eq:z}
 z_{l,n}[i]= y_{l,n}[i+d_l]
 \end{align}
 where $J$ is the aperture size, and $d_l$ denotes the specific detector offset to indicate the active channel elements, which is determined
 for each scan line index $l$. 
% for the conversion between the data cube $\Xb$ and $\Zb$,
% where 
 This implies that
 the dark triangular  regions in $\Xb$ as shown Fig.~\ref{fig:data_processing_pipeline}(a), which correspond to inactive receiver elements, are removed in constructing $\Zb$.

  The standard  delay and sum (DAS) beamformer for the $l$-th scanline at the depth sample $n$ can be expressed as
\begin{equation}\label{eq:DAS}
{u}_{l,n} 
=\frac{1}{J}\mathbf{1}^\top\zb_{l,n} %,  \quad l=0,\cdots, L-1,
\end{equation}
where $\mathbf{1}$ denotes a $J$-dimensional column-vector of ones. %, and $J$ is the number of active {channels}.
On the other hand, DeepBF \cite{khan2019deep,khan2020adaptive} generates multiple scan-lines in a depth  simultaneously
using  multiple-depth channel data:
\begin{eqnarray}\label{eq:Oc}
\ob_n
~= \Gc_\Thetab\left(\Zb_n\right),&& \ob_n \in \Oc
\end{eqnarray}
where $ \ob_n$ denotes the output at the depth $n$:
 \begin{align}\label{eq:on}
 \ob_n= \begin{bmatrix} o_{0,n} \\ \vdots \\ o_{L-1,n}  \end{bmatrix} 
\end{align}
with  $o_{l,n}$ referring to  the target image at the $l$-th scan line and the depth $n$, and
 $\Gc_\Thetab$ is a neural network parameterized by $\Thetab$,
 $\Zb_n$ is a channel data collected from, for example,  seven depth planes around the depth $n$:
 \begin{align}\label{eq:Zb}
 \Zb_n = \begin{bmatrix}\zb_{n-3} & \cdots  & \zb_n & \cdots & \zb_{n+3} \end{bmatrix} \in \Zc
 \end{align}
 %with $\Zc$ referring the input space,
where  $\zb_n$  is defined as:
 \begin{align}\label{eq:zbn}
 \zb_n = \begin{bmatrix} z_{0,n} \\ \vdots \\ z_{L-1,n}  \end{bmatrix} % &\quad \quad \ob_n= \begin{bmatrix} o_{0,n} \\ \vdots \\ o_{L-1,n}  \end{bmatrix} 
\end{align}
Then, the resulting neural network training is given by
 \begin{eqnarray}\label{eq:train_new}
\min_{\Thetab} \sum_{t,n}\|\ob_n^{(t)} - \Gc_\Thetab (\Zb_n^{(t)}) \|_2^2,
\end{eqnarray}
where $(\Zb_n^{(t)},\ob_n^{(t)})$ denotes the 
input channel data and the target image data at the depth index $n$,
and the superscript index $t$ is the training data set index.
In fact, the training data can be obtained not only from different experiment data but also from different depth in the same frame \cite{khan2019deep,khan2020adaptive}.

One of the important advantages of the deep beamformer is that
depending on the target data $\ob_n$, the neural network parameter $\Thetab$ is trained
accordingly to mimic target data.  In \cite{khan2019deep,khan2020adaptive}, the authors
used the DAS, minimum variance beamformer (MVBF), and deconvolution images as targets,
so that deep beamformer can directly produce target domain images.
However, as discussed before, separate network parameter $\Thetab$ should be stored
for each target domain, demanding significant amount of scanner resources.

\subsection{Adaptive Instance Normalization (AdaIN)}

AdaIN \cite{huang2017arbitrary} is an extension of instance normalization \cite{ulyanov2016instance}, 
which goes beyond the classical role of the normalization methods for image style transfer.
The key idea of AdaIN is that latent space exploration is possible 
by adjusting the mean and variance of the feature map.
Specifically, style transfer is performed by matching the mean and variance of the feature map of the input image to those of the style  image \cite{huang2017arbitrary}.

More specifically, let a multi-channel feature map be represented by
\begin{align}\label{eq:X}
\Ub =&\begin{bmatrix} \ub_1  & \cdots &\ub_P \end{bmatrix} \in \Rd^{HW\times P},
\end{align}
where 
$\ub_i  \in \Rd^{HW\times 1}$ refers to the $i$-th column vector of $\Ub$, 
which represents the vectorized feature map of size of $H\times W$  at the $i$-th channel.
Suppose, furthermore, the corresponding feature map for the style reference image is given by
 \begin{align}\label{eq:Y}
\Vb =&\begin{bmatrix} \vb_1  & \cdots &\vb_P \end{bmatrix} \in \Rd^{HW\times P} .
\end{align}
Then, AdaIN changes the feature data for each channel using the following transform:
\begin{align}
\zb_i &= \Tc(\ub_i,\vb_i),\quad i=1,\cdots, P
\end{align}
where
\begin{align}\label{eq:AdaIN}
\Tc(\ub,\vb)  := \frac{\sigma(\vb) }{\sigma(\ub) }\left(\ub -\mu(\ub) \1\right) +\mu(\vb)\1,\quad 
\end{align}
where $\1 \in \Rd^{HW}$ is the $HW$-dimensional vector composed of 1, and 
 $\mu(\ub)$ and $\sigma(\ub)$ are the mean and standard deviation (std) for $\ub\in \Rd^{HW}$, which are computed by:
\begin{align}
\mu(\ub)  =  \frac{1}{HW}\1^\top \ub ,~&
\sigma(\ub)  = \sqrt{ \frac{1}{HW} \| \ub  -\mu \1\|^2}
\end{align}

{The AdaIN transform in \eqref{eq:AdaIN}  has an important link to optimal transport \cite{peyre2019computational,villani2008optimal}.
Specifically,  consider 
 the optimal transport scheme between two probability
spaces $\mathcal{U} \subset \Rd^{HW}$ and $\mathcal{V} \subset \Rd^{HW}$, 
 equipped with the Gaussian probability measure $\mu\sim \mathcal{N}(\mb_U,\Sigmab_U)$  and
$\nu\sim\mathcal{N}(\mb_V,\Sigmab_V)$, respectively, where $\mb_U$ and $\Sigmab_U$ denote the mean
vector and the covariance matrix, respectively.
Then, a recent  study \cite{mroueh2019wasserstein} showed that the
optimal transport map from the measure $\mu$ to the measure $\nu$ with respect to Wasserstein-2 distance is given by
\begin{align}\label{eq:Tmap}
T_{\mu\rightarrow \nu}(\ub)= \mb_V + \Sigmab_U^{-\frac{1}{2}}\left(\Sigmab_U^{\frac{1}{2}}\Sigmab_V\Sigmab_U^{\frac{1}{2}}\right)^{\frac{1}{2}}\Sigmab_U^{-\frac{1}{2}}(\ub-\mb_U)
\end{align}
Therefore, if we assume the i.i.d. distribution, i.e.
\begin{align}
\mb_U=\mu(\ub)\1, \Sigmab_U=\sigma(\ub)\Ib, & & \mb_V=\mu(\vb), \Sigmab_V=\sigma(\vb)\Ib
\end{align}
where $\Ib$ is the identity matrix, then the optimal transport map in \eqref{eq:Tmap}
can be simplified to
\begin{align}\label{eq:T}
T_{\mu\rightarrow \nu}(\ub)= \mu(\vb)\1 + \frac{\sigma(\vb)}{\sigma(\ub)}(\ub-\mu(\ub)\1)
\end{align}
which is equivalent to the AdaIN transform in \eqref{eq:AdaIN}.
Therefore, the AdaIN transform can be considered  the minimal cost transport path when converting one feature map to another
under i.i.d. Gaussian feature map assumption.
}

\section{Theory}
\label{sec:theory}

\subsection{Switchable Deep Beamformer using AdaIN Layers}

Inspired by the observation that AdaIN is an optimal transport plan between the feature layers,
our goal is to implement multi-purpose deep beamformer using a single baseline network followed by optimal transport layers to specific target distributions.
Specifically, we use DeepBF \cite{khan2019deep,khan2020adaptive}
as  the baseline network
and then use AdaIN transform to transport the features to different domain features. % as shown in Fig.~\ref{fig_CycleGAN}(b).

More specifically,  our switchable deep beamformer is formally defined as
\begin{align}
\Gc(\cdot;\cdot): \Zc \times \Wc \mapsto \Oc
\end{align}
where $\Zc$ denotes the input RF channel data space composed of elements in \eqref{eq:Zb},
$\Oc$ is the output space for the beamformed data in \eqref{eq:Oc},
and $\Wc$ is the AdaIN code vector space, whose element $\wb_c\in \Wc$ is
generated
by the AdaIN code generator $\Fc: \Rd \mapsto \Wc$ as
\begin{align}
\wb_c = \Fc(c)
\end{align}
where $c\in \Rd$ is the target style code.
AdaIN code generator takes the style code representing a target domain,
 and generates the style mean and variance vectors required for that domain.

The architecture of the proposed switchable network $\Gc$ is shown in  Fig.~\ref{fig:network_architecture}.
%where $\Thetab$ is the network parameters,
%$\sb_n$ is the network input in \eqref{eq:sb}, and
%$\Fc_\Psib(c)$ refers to the AdaIN code generator parameterized by $\Psib$,
%which  generates AdaIN code for a given style code $c$. 
In the baseline network, there are nine convolution blocks shown in blue and yellow colors. The blue blocks are composed of two convolution layer with ReLU activation function, while yellow block is composed of convolution, and Leaky ReLU blocks. All convolution layers uses stride value of $(1,1)$, with a 2-dimensional filter that has a dimension of $3\times 3$, expect the last layer shown in green color which uses $(7,1)$, with a 2-dimensional filter that has a dimension of $7\times 1$. 
As  shown in  Fig.~\ref{fig:network_architecture}, there exists an AdaIN layer at the bottleneck layer.
The AdaIN layers take a mean vector and a variance vector as input.
During the training phase, the AdaIN layers take the output vectors from the AdaIN code generator $\Fc$ in Fig.~\ref{fig:network_architecture},
 which consists of fully connected layers and is trained together with the baseline neural network.  
For mean vector generation, a linear activation was used, whereas for the variance generation ReLU activation function was used to impose nonnegativity.

At the inference phase, we only use the baseline network $\Gc$ with the trained AdaIN codes $\{\wb_c\}$, 
as the AdaIN code generator is no more necessary.
Then, depending on the specific AdaIN code $\wb_c$, our switchable deep beamformer output 
$$\ob_n(c) = \Gc(\Zb_n;\wb_c)$$
becomes an image with the $c$-style processing.
Accordingly,  the total memory requirement can be significantly reduced,
and multiple processing can be done instanteneously by simply changing the code $\wb_c$.

{\subsection{Target Image Generations}}
\label{sec:target_output}
{As for multiple domain target reference data, in addition to DAS,
we consider images from
three post processing  methods: (1) deconvolution (deblurring),  (2) despeckle (denoising), and (3) the combination
of the two.} More details are as follows.

\subsubsection{DAS}

In contrast to \cite{khan2019deep,khan2020adaptive} where IQ signal was used as target, for proposed switchable deep beamformer, we use the
log-compressed absolute value of Hilbert transformed RF-sum as  DAS target.
Thanks to such changes, all the beamforming process is now learned by the neural network,
and only thresholding for specific dynamic range can be performed for visualization as shown in  {Fig}.~\ref{fig:data_processing_pipeline}(c).
Similar log-compressed domain learning approaches are taken for other processing as follows.

\subsubsection{Deblurring}

\begin{figure}[!hbt]
\centerline{\includegraphics[width=7cm]{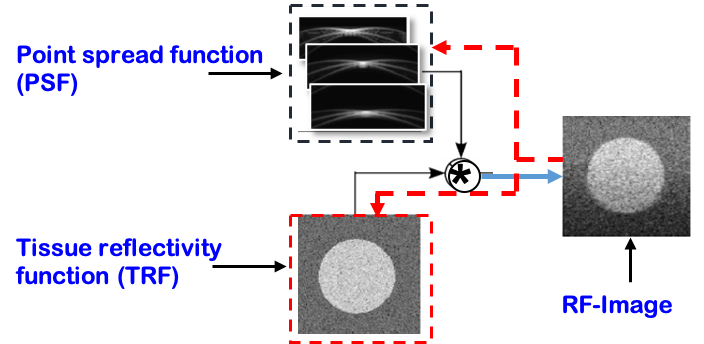}}
\caption{Block diagram of deconvolution ultrasound imaging.}
\label{fig:Deconv_model}
\end{figure}

{Deconvolution method can help reduce the blur to improve spatial resolution of an ultrasonic imaging system.  As shown in Fig.~\ref{fig:Deconv_model}, in US, a received signal is modeled as a convolution product of tissue reflectivity function (TRF) and point spread function (PSF), where TRF represents scatter's acoustic properties, while the impulse response of the imaging system is modeled by PSF. 
The PSF is determined by the detector configurations, focused beam width, etc.}

%For the deconvolution US,  the forward model in \eqref{eq:fwd} of the received signal can be solved by basis pursuit method \cite{chen2001atomic} and

Then, the unknown TRF $x$ is estimated by solving the following $l_1$ minimization problem: % it is expressed as :
\begin{equation}
\label{equ:Deconv_model}
\hat x = \argmin{x} ||y - h \ast x||^{2} + \lambda_{1}||x||_{1}
\end{equation}
where $y$  is the Hilbert transformed DAS RF-sum, 
 $h$ is the PSF, $\lambda_{1}=0.02$ is a regularization parameter empirically chosen to control the tradeoff between sparsity and fitting fidelity.

%
%\new{
%\begin{eqnarray}\label{eq:fwd}
%\ub&=&h \ast \db + \zeta \ ,
%\end{eqnarray}
%}
%where $\ub$, is the Hilbert transformed DAS RF-sum, $\zeta$ is the  noise, and $h$ is the PSF. 

In most practical cases, the complete knowledge of $h$ is not available, and therefore both unknown TRF $x$ and  the PSF $h$ have to be estimated together, which is called the blind deconvolution problem. In this paper, we first estimate $h$ from $y$, and then $x$ is estimated based on $h$ \cite{7565583}. To train the proposed switchable beamformer, a log-compressed absolute value,  $20\log_{10}(|\hat x|)$, is used as a training reference.

\subsubsection{Denoising}

\begin{figure}[!hbt]
	\centerline{\includegraphics[width=9cm]{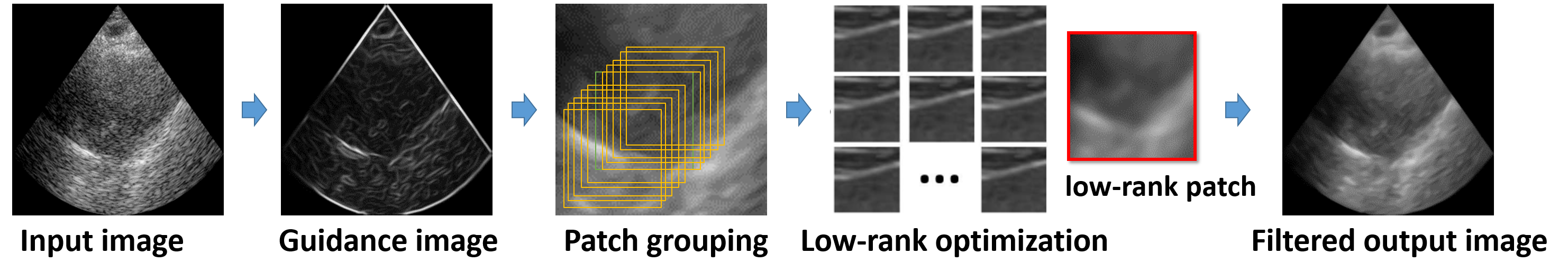}}
	\caption{Block diagram of non-local low-rank (NLLR) filtering-based speckle removal method.}
	\label{fig:NNLR_model}
\end{figure}

{Speckle noises are the granular patterns appears in US images, which is one of the a major reasons of quality degradation and removal of it can subsequently enhance the structural details in US images \cite{NLLR_Despeckle}.}
{To design a speckle denoising beamformer, the log-absolute value of Hilbert transformed RF-sum %  $20\log_{10}(|\hat y|)$ 
was filtered using non-local low-rank (NLLR) method \cite{NLLR_Despeckle} to generate the reference data for supervised training. In NLLR as shown in Fig.~\ref{fig:NNLR_model}, the image is pre-processed to generate a guidance map and later non-local filtering operations are performed on the candidate patches that are selected using that guidance map.  For further refinement of filtered patches, a truncated nuclear norm (TWNN) and structured sparsity criterion are used \cite{zhang2012matrix, gu2014weighted}.}

\subsubsection{Deconvolution + Denoising}

To generate the deconvolution and despeckle reference images,
we applied the NLLR filter on deconvoluted log-absolute RF-sum to generate deconvolution+despeckle output.

%
%{Mathematically $$\ob = \Fc(20log_{10}(|\ub|);\beta;\kb)$$, where $\Fc$ NLLR-based filter, $\beta$ is a patch-size which was set to $(7\times7)$, $\kb=2$ is the number of iterations.} 

\subsection{Switchable Deep Beamformer Training}

Let the data set  $\{(\Zb_n^{(t)},\ob_n^{(t)}(c))\}_{c,n,t}$ denotes the training data set collected across all depth planes $n$ and style $c$,
where $\Zb_n^{(t)}$ is defined in \eqref{eq:Zb} and $\ob_n(c)$ is the target image for a given style $c$ at the depth $n$.
Then, the resulting neural network training for the proposed switchable deep beamformer is given by
 \begin{eqnarray}\label{eq:train_AdaIN}
\min_{\Gc,\Fc}\sum_{c,t,n} \|\ob_n^{(t)}(c) - \Gc (\Zb_n^{(t)}; \Fc(c)) \|_2^2,
\end{eqnarray}
For each target style $c$, the AdaIN code generator network $\Fc$ expands the style input to produce desired size AdaIN code $\wb_c$ with the help of multiple fully-connected layers. In particular, the style code is given by  $c\in \{-1, -0.5, 0.5,-1\}$  to generate four different AdaIN code vectors representing DAS, Despeckle, Deconvolution, and Deconvolution+Despeckle targets, respectively.
Once the neural network is trained and  AdaIN codes $\{\wb_c\}$ are learned,  our switchable deep beamformer  can generate multiple domain outputs by simply changing the AdaIN code.

\section{Method}
\label{sec:methods}

\subsection{Dataset}
Three dataset were collected using E-CUBE 12R US system with  L3-12H linear array probe. The first dataset was acquired with an operating frequency of 8.5MHz from the carotid/thyroid regions of $10$ volunteers, in which $40$ frames were scanned from four different regions (carotid artery, trachea, thyroid right and left lobes) of each subject providing a set of $400$ images. The second dataset was acquired with an operating of 10MHz from the calf and forearm region of a volunteers,  where $50$ frames were acquired from each region resulting in a set of $100$ frames.  The last dataset was generated using ATF-539 phantom with an operating frequency of 8.5MHz, 10MHz and 11.5MHz, different regions of phantom were scanned and in total $218$ frames were recorded. The length of the lateral axis for all samples was $38.4mm$, the axial depth (AD) was varied from $20$ to $80mm$ and the focal depth was also adjusted accordingly. For all samples $64$ active channels (CH) were used to acquire $96$ scan-lines (SC/L) using standard single-line-acquisition (SLA) method.

\subsection{Network training}

The input and output data configurations are shown in {Fig}.~\ref{fig:data_processing_pipeline}(a) and (b), respectively. Similar to the DeepBF\cite{khan2019deep,khan2020adaptive}, herein we  use a multi-line, multi-depth approach to form a neural network input. First, a raw RF-signal cube was acquired which is later time-delayed to correct the time-of-flight in out-of-phase signals. The RF data cube depends on the axial depth of the image, for this study, we used variable depths ranges from $20$ to $80mm$ axial depth planes representing $1280\sim 4608$ axial samples.  From a time corrected cube a subset of size $7\times 96\times64$ is selected  to produce $1\times 96\times1$ output. 

To train our model using \eqref{eq:train_AdaIN}, input output pairs composed of channels data cube and targeted depth plane are collected. Specifically,
the model was trained using a set of $148,204$ cubes, which are randomly selected from $8$ frames of a subject, which are divided into $140,794$ samples for training and $7,410$ samples for validation. The remaining dataset of $382$ carotid/thyroid and $218$ phantom were used as a test data. In addition, to see the generalization capability of the algorithm, $100$ frame data from totally different anatomical regions (forearm/calf muscles) were used as an independent dataset.

The network was implemented with Python3, using Keras and TensorFlow platform. Specifically, for network training, the parameters were estimated by minimizing the $l_2$ norm loss function using  Adam stochastic gradient descent. The learning rate started from $10^{-4}$ and {exponentially} decreased with the $10^{3}$ decay steps at a decay rate of $0.999$ for $100$ epochs. The training was stopped after $69$ epochs using early-stopping method with patience steps of $20$.

\subsection{Comparison methods}
For the comparative studies of our switchable deep beamformer method,  we use standard DAS, and DeepBF method \cite{khan2019deep,khan2020adaptive}. The DAS is a simple sum of channel data
as described in \eqref{eq:DAS}.
DeepBF \cite{khan2019deep,khan2020adaptive} was trained to produce deconvolution (Deconv.BF).

{For the DeepBF training, the target data was generated as defined in Section \ref{sec:target_output}. Default configuration of the DeepBF in  \cite{khan2019deep,khan2020adaptive} was used, where $9$ convolution blocks each consisting of a batch-normalization layer, $4$ convolution layers with ReLU activation function and $64$ filters of $(3,3)$ kernel-size was used. The input cube was $64\times96\times3$ and target output was $1\times96\times2$.}

\subsection{Performance metrics}

To evaluate the performance of our proposed beamforming method, we used three measures of contrast namely contrast-recovery (CR), contrast-to-noise ratio (CNR) and generalized contrast-to-noise ratio (GCNR). 

The contrast in ultrasound images can be calculate with the help of two regions of interest defined as target ($R_t$) and background ($R_b$). Typically binary masks are used to select two regions in an image, and  in this study we prepare two separate masks for each image,  and the selected regions are highlighted in the B-mode results.

The CR measure is the most common measure of contrast; however, it only considers the mean intensity difference between two regions. Since ultrasound images are naturally prone to noise (especially the speckle noise) which can degrade the visual quality, so a better measure of contrast is CNR. Mathematically the CR and CNR are defined as follows

\begin{equation}
 {\hbox{CR}}(R_t,R_b) = |\mu_{R_t}-\mu_{R_b}|
 \end{equation}
 \begin{equation}
 {\hbox{CNR}}(R_t,R_b) = \frac{|\mu_{R_t}-\mu_{R_b}|}{\sqrt{\sigma^2_{R_t} + \sigma^2_{R_b}}},
 \end{equation}
where $\mu$ and $\sigma$ are the local means, and the standard deviations of the target ($R_t$) and background ($R_b$) \cite{BiomedicalImageAnalysis}, respectively. 

The CNR works well for most of the cases; however, it is scale-variant and sensitive to the dynamic range, which make it difficult to interpret. Recently a better generalization of contrast statistics was proposed called generalized CNR (GCNR) \cite{rodriguez2019generalized}. The GCNR measure is based on the overlap between two distributions and mathematically it is defined as
 
\begin{equation}
 {\hbox{GCNR}}(R_t,R_b) = 1- \int \min \{p_{R_t} (v), p_{R_b} (v) \} dv,
 \end{equation}
 where $p_{R_t}$, and $p_{R_t}$ are the probability distributions of target and background pixel value $v$, respectively. In ideal case, target and background must be statistically independent and there should not be any overlap, resulting in perfect score GCNR=1, while in the worst case when they completely overlap, the GCNR will be $0$.

\section{Experimental Results}
\label{sec:results}
In this study we  verified our switchable deep beamformer using three different dataset mentioned in Section \ref{sec:methods} and compared the results with DAS, and DeepBF \cite{khan2019deep,khan2020adaptive}. All the figures are scaled in the dynamic range of $-60\sim0$dB. 
	
\subsection{Qualitative Comparison}
\subsubsection{Tissue mimicking phantom}
\begin{figure*}[!hbt]
	%	\centerin
	\centerline{\epsfig{figure=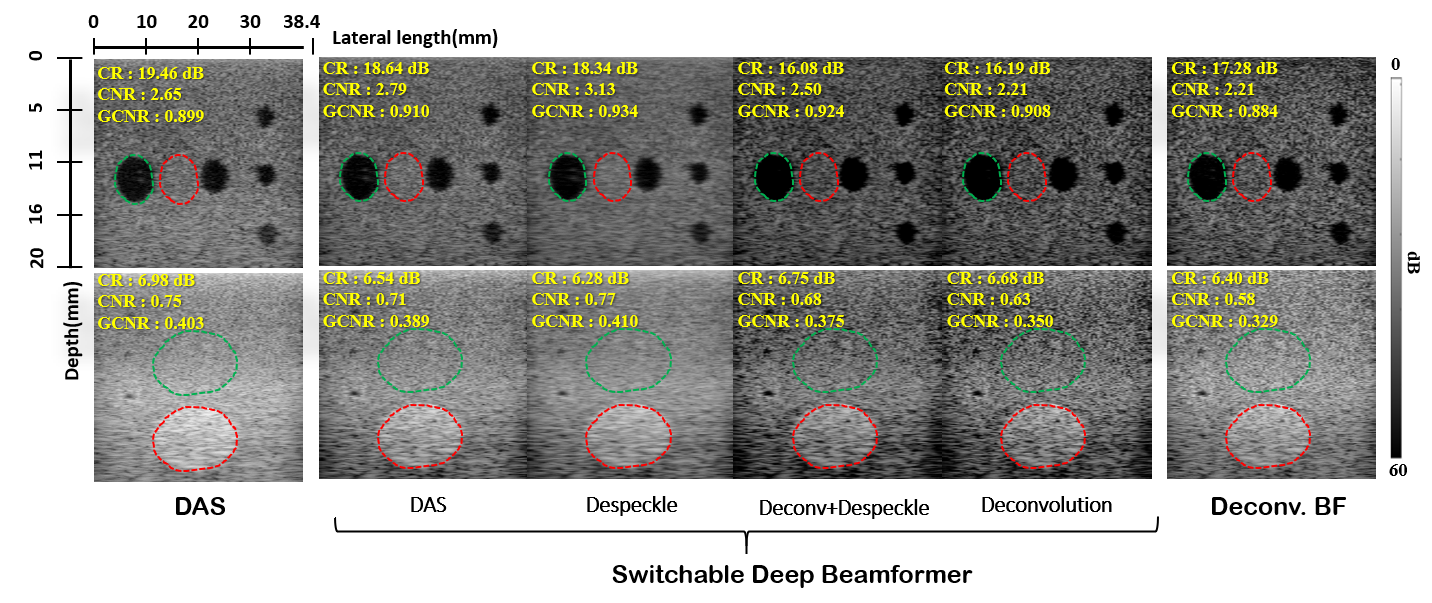, width=18cm}}   
	\caption{\footnotesize Reconstruction results of standard DAS-BF, Deconv.BF, and the proposed switchable deep beamformer for tissue mimicking phatom data from anechoic and hyperechoic regions.}
	\label{fig:results_view_invivo_full}	
\end{figure*}

To validate the reconstruction performance of the proposed switchable deep beamformer method, in Fig.~\ref{fig:results_view_invivo_full}, we showed the results of two phantom images generated from the anechoic and hyper-echoic regions. From the results it can be seen that the proposed model can generate four different types of output depending on the style code. Interestingly, in all cases the visual quality is either comparable or even better than the comparative algorithms. For the given image, two regions of interests shown in red and green colors were selected to calculate the CR, CNR and GCNR statistics. 

With a careful observation, we can see that there is speckle noise in DAS, and switchable deep beamformer DAS output, which is further enhanced by deblurring filtration. The deblurring filtration improves the sharpness of the image; however, it also increases the noise, which results in reduced CNR and GNCR. Although the contrast statistics of deconvolution beamformers is not good but the image looks much sharpers. Especially in hyperechoic case where standard methods shows blurry images with washout artifacts. 

The speckle noise can be reduced using switchable deep beamformer (Despeckle): as it can be seen in contrast statistics and example image in Fig.~\ref{fig:results_view_invivo_full}, the speckle patterns in the background and target are reduced. To obtain noise-free sharp image the switchable deep beamformer can be switched to (Deconv + Despeckle), which makes the output image notably sharper with reduced speckle noise.

\subsubsection{\textit{In-vivo} test data}
\begin{figure*}[!hbt]
	%	\centerin
	\centerline{\epsfig{figure=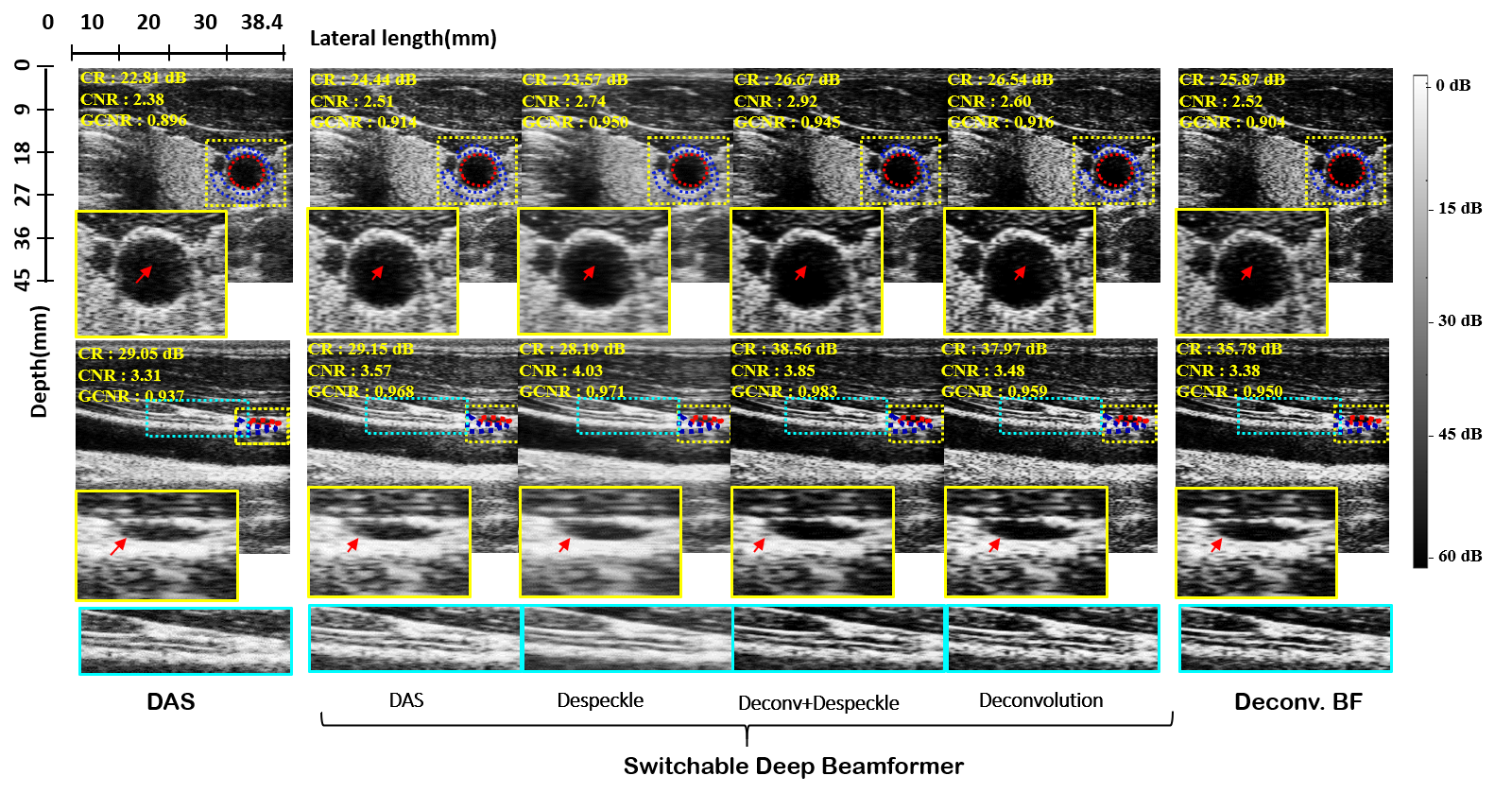, width=18cm}}   
	\caption{\footnotesize Reconstruction results of standard DAS-BF, Deconv.BF, and the proposed switchable deep beamformer for \textit{in-vivo} scans from carotid and artery regions.}
	\label{fig:Bmode_test_carotid_artery}	
\end{figure*}

In Fig.~\ref{fig:Bmode_test_carotid_artery}, we compared two \textit{in-vivo} examples scanned from the carotid and artery regions using $8.5$ MHz center frequency. Here again the proposed switchable deep beamformer achieves better performance compared to DAS and Deconv.BF methods especially in (Deconvolution/DeSpeckle) case which looks much sharper and have relatively less noise. 

\begin{figure*}[!hbt]
	%	\centerin
	\centerline{\epsfig{figure=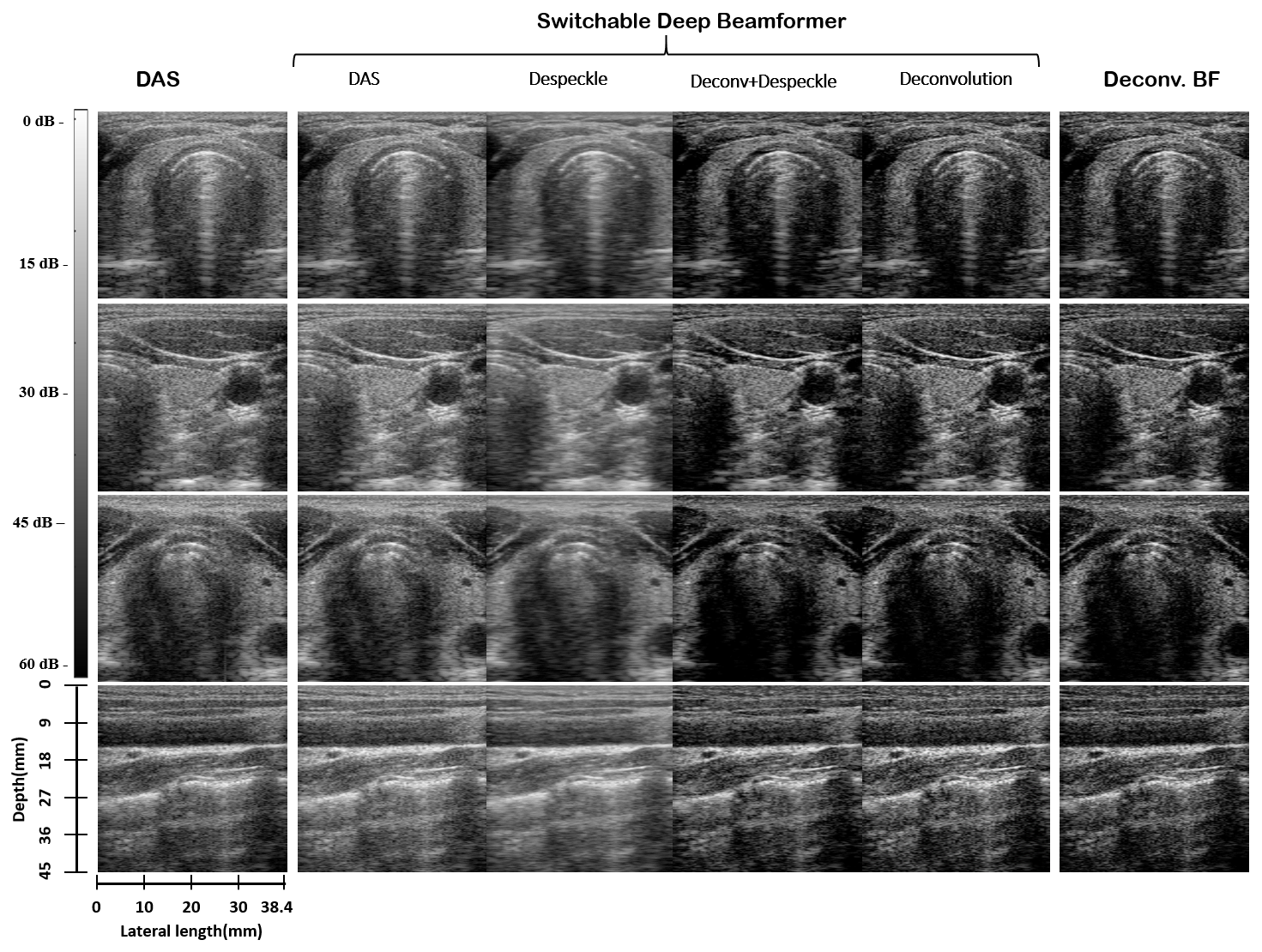, width=18cm}}   
	\caption{\footnotesize Reconstruction results of standard DAS-BF, Deconv.BF, and the proposed switchable deep beamformer for \textit{in-vivo} scans from trachea, carotid, thyroid (central lobe), and artery region}
	\label{fig:Bmode_test_invivo_phantom}	
\end{figure*}

To further test the reconstruction quality of switchable deep beamformer, we showed four additional results  in Fig.~\ref{fig:Bmode_test_invivo_phantom}. From the figures it can be easily seen that the switchable deep beamformer can perform different tasks such as simple image formation, deconvolution, speckle suppression, and noise-suppressed deconvolution effectively.

\subsubsection{\textit{In-vivo} independent data}
\begin{figure*}[!hbt]
	%	\centerin
	\centerline{\epsfig{figure=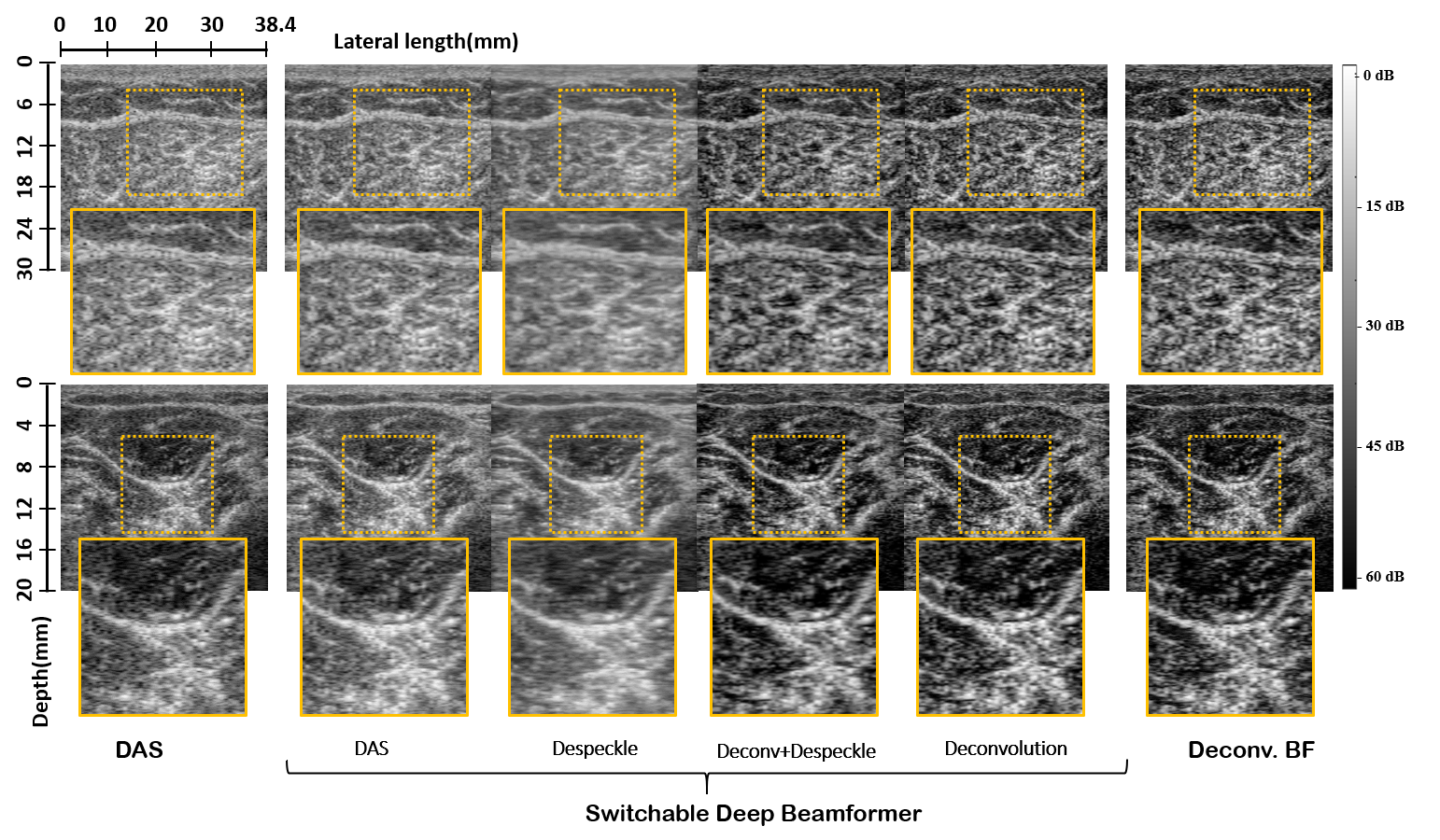, width=18cm}}   
	\caption{\footnotesize Reconstruction results of standard DAS-BF, Deconv.BF, and the proposed switchable deep beamformer for \textit{in-vivo} scans from calf and forearm regions.}
	\label{fig:Bmode_ind_forearm_calf}	
\end{figure*}

In this experiment we used a dataset acquired from two completely different body parts scanned using $10$MHz operating frequency. In Fig.~\ref{fig:Bmode_ind_forearm_calf}, we showed the results of two \textit{in vivo} examples from calf and forearm regions.  The images are generated using standard DAS, Deconv.BF,  and the proposed switchable deep beamformer method.   In Fig.~\ref{fig:Bmode_ind_forearm_calf}, it can be easily seen that our method perform all four tasks effectively using a single model. In despeckle beamformer case, the switchable deep beamformer preserves the structural details while notably reduces the noise level. Interestingly, with combined deconvolution and despeckle, we can achieve two tasks (deblurring and denoising) tasks simultaneously. 

It is also noteworthy to point-out that the switchable deep beamformer was trained using only a subset of RF-data acquired from $8$ frames of a carotid/thyroid regions of a subject scanned with $8.5$MHz operating frequency. However, the performance in very diverse test scenarios is still remarkable, which clearly shows the generalization power of the proposed method.

\subsection{Quantitative comparison}
To quantify the performance gain we  compared the CR, CNR, and GCNR statistics of reconstructed B-mode images for all three datasets. Quantitatively CR, CNR, and GCNR values of switchable deep beamformer were slightly improved compared to the existing methods.  Table~\ref{tbl:results_STATS} shows the comparison of DAS, Deconv.BF and proposed switchable deep beamformer method on all three dataset. In terms of CR, CNR and GCNR, the overall performance of switchable deep beamformer is remarkable. In carotid, thyroid, artery \& trachea dataset,   switchable deep beamformer for deconvolution
produce highest CR among all while maintaining the noise level lower than the Deconv.BF. A similar trend can be seen in other dataset.

As expected the CNR and GCNR statistics of despeckle style switchable deep beamformer are highest as these measures are based on the sum of noise and background noise variance. Since despeckle style switchable beamformer effectively reduces the noise, it produces best statistics for noise based contrast measures. One interesting feature of the proposed method is its simultaneous deblurring and denoising capability. In Deconvolution+Despeckle mode, our method performs two competing tasks of deblurring and denoising, and surprisingly it enhances the CR statistics while maintaining the noise levels. 

Interestingly, the proposed switchable beamformer also enhances the quality of B-mode images in DAS mode. In particular, for dataset $a$, $b$, and $c$ the GCNR values of proposed method in DAS mode are $0.02$, $0.011$, and $0.005$ units higher than the conventional DAS, respectively. Similarly, the deconvolution mode of switchable beamformer also outperform the Deconv.BF method. In particular, for dataset $a$ and $b$ it produced $0.016$, and $0.013$ units higher GCNR. While for dataset $c$ the results are quite comparable.

Since the encoder of the proposed switching beamformer was trained to produce a common representation for all types of target output, while the switching function was performed on decoder side, one possible reason for the enhanced performance would be that the network can learn better representation of the channel data through separate paths of forward and skip connections for contrast and speckle patterns, respectively. This may help the decoder to filter particular representation combination depending on the AdaIN encoded output. 

{\begin{table*}[!hbt]
	\centering
	\caption{Comparison of performance statistics with different beamformers}
	\label{tbl:results_STATS}
	\resizebox{0.9\textwidth}{!}{
		\begin{tabular}{c|c|ccc|ccc|ccc}
			\hline
			\multicolumn{2}{c|}{{Beamforming}} & \multicolumn{3}{c|}{{CR (dB)}} & \multicolumn{3}{c|}{{CNR}} & \multicolumn{3}{c}{{GCNR}}  \\
			\multicolumn{2}{c|}{{Method}} & \textit{a} & \textit{b} & \textit{c} & \textit{a} & \textit{b} & \textit{c} & \textit{a} & \textit{b} & \textit{c}  \\ \hline\hline
			\multicolumn{2}{c|}{DAS} & 18.53 & 16.97 & 24.92 & 2.29 & 2.17 & 2.63 & 0.824 & 0.785 & 0.923 \\
			\hline
			\multirow{4}{*}{Switchable DeepBF} & (DAS) & 18.81 & 15.77 & 24.71 & 2.45 & 2.30 & 2.63 & 0.844 & 0.796 & 0.928 \\
			 & (Despeckle) & 18.13 & 15.49 & 23.95 & 2.92 & 2.59 & 3.02 & 0.888 & 0.821 & 0.958 \\
			 & (Deconvolution) & 20.13 & 13.70 & 26.79 & 2.00 & 1.82 & 2.20 & 0.773 & 0.790 & 0.858 \\
			 & (Deconvolution + Despeckle) & 20.02 & 13.63 & 26.31 & 2.29 & 2.04 & 2.43 & 0.822 & 0.811 & 0.899 \\
			 \hline
			 \multicolumn{2}{c|}{Deconv.BF} & 19.78 & 17.27 & 27.55 & 1.93 & 1.92 & 2.37 & 0.757 & 0.777 & 0.884 \\\hline
			 
		\end{tabular}
	}
	\\
	\small{$^a$ Test: \textit{carotid, thyroid, artery \& trachea}, $^b$ Test: \textit{tissue mimicking phantom}, $^c$ Independent: \textit{calf \& forearm}}
	\vspace*{-0.3cm}
\end{table*}
}

 \subsection{Computational time}
 
One big advantage of proposed switchable deep beamformer method is its reduced complexity, which allows for easily adaptation as only one model is used for various reconstruction tasks. This also allows fast reconstruction of ultrasound images with desired post processing filtration effects.  The average reconstruction time for each depth planes is around $0.165$ (milliseconds), which is significantly lower than the conventional beamformer+filtration (deblurring/despeckle) methods. Interestingly, with combined learning of multiple tasks through AdaIN, the proposed model can achieve DAS and Deconv.BF like results with relatively shallower model compared to DeepBF which took around $4.8$ (milliseconds). The implementation method used in current study can be optimized by replacing the fully-connected neural network with fixed style vector values and optimized implementation for parallel reconstruction of multiple depth planes.

\section{Conclusion}
\label{sec:conclusion}
In this paper, we presented a novel switchable deep beamformer
to generate desired quality B-mode ultrasound image directly from channel data with different post filtration effects  by simply changing the AdaIN code.
The proposed method is purely a data-driven method which harnesses the information in the spatio-temporal dimensions along the scan-line, axial depth and active channels measurement, which help in generating improved quality B-mode images compared to standard beamforming methods. 
In the proposed method, a single network was trained to produce four types of output including DAS, deconvolution, despeckle and deconvolution + despeckle directly from the channel data.
The proposed method improved the contrast of B-modes images by preserving the dynamic range and structural details of the RF signal in both the phantom and \textit{in vivo} scans, and can effectively perform computationally expensive post filtration task during the beamforming process. This post training adaptivity through AdaIN codes allow easily implementation; and therefore, this method can be an important platform for data-driven ultrasound beamformer. 

%\bibliographystyle{IEEEtran}
%\bibliography{main}


\end{document}

%% file: packages.tex
\usepackage{multirow}
\usepackage{tabu}
\usepackage{outline}
\usepackage{pmgraph}
\usepackage[normalem]{ulem}
\usepackage[utf8]{inputenc}
\usepackage{amssymb}
\usepackage{hyperref}
\usepackage{amsmath}
\usepackage{graphicx}
\usepackage{times}
\usepackage{xcolor}
\usepackage{xspace}
\usepackage[colorinlistoftodos]{todonotes} % for \todo
\usepackage{cite}
\usepackage{bm} % bold
\usepackage{url}

\usepackage{epstopdf}
\AppendGraphicsExtensions{.tiff}
\graphicspath{{fig/}} % Directory in which figures are stored

\usepackage{epsfig}
\usepackage{tikz}
\usetikzlibrary{spy}
\usepackage{algpseudocode}
\usepackage{algorithm}
\usepackage{mathrsfs}

%% file: macros.tex
\long\def\comment#1{} % comment out text

\DeclareMathOperator*{\argminop}{arg\,min} % thin space, limits underneath
\def\argmin#1{\argminop_{#1}}
\newcommand{\xmath}[1] {\ensuremath{#1}\xspace}
\newcommand{\blmath}[1] {\xmath{\bm{#1}}}

\newcommand{\1}{\blmath{1}}

\newcommand{\Ib}{{\blmath I}}

\newcommand{\Ub}{{\blmath U}}
\newcommand{\Vb}{{\blmath V}}

\newcommand{\Xb}{{\blmath X}}
\newcommand{\Yb}{{\blmath Y}}
\newcommand{\Zb}{{\blmath Z}}

\newcommand{\mb}{{\blmath m}}

\newcommand{\ob}{{\blmath o}}

\newcommand{\ub}{{\blmath u}}
\newcommand{\vb}{{\blmath v}}
\newcommand{\wb}{{\blmath w}}

\newcommand{\yb}{{\blmath y}}
\newcommand{\zb}{{\blmath z}}

\newcommand{\Tc}{\mathcal{T}}
\newcommand{\Gc}{\mathcal{G}}

\newcommand{\Zc}{\mathcal{Z}}
\newcommand{\Oc}{\mathcal{O}}

\newcommand{\Wc}{\mathcal{W}}

\newcommand{\Thetab}{{\boldsymbol {\Theta}}}
\newcommand{\Sigmab}{{\boldsymbol {\Sigma}}}
\newcommand{\Rd}{{\mathbb R}}

\newcommand{\beq}{\begin{equation}}
\newcommand{\eeq}{\end{equation}}
\newcommand{\beqa}{\begin{eqnarray}}
\newcommand{\eeqa}{\end{eqnarray}}
\newcommand{\Fc}{{\mathcal F}}